\documentclass[apj]{emulateapj}
\usepackage{amsmath,amssymb}
\usepackage{epsfig}
\usepackage{natbib}
\usepackage{url}
 \usepackage{hyperref}
\usepackage{subfigure}
\usepackage{booktabs}

\received{\today}
\accepted{}

\shorttitle{An ultra-diffuse galaxy behind M31}
\shortauthors{Mart\'{\i}nez-Delgado et al.}

\newcommand{\DGSAT}{DGSAT}
\newcommand{\DGSATI}{\DGSAT~I}
\newcommand{\etal}{{et al.~}}

\newcommand{\cm}{\,\mathrm{cm}}
\newcommand{\kpc}{\,\mathrm{kpc}}
\newcommand{\Mpc}{\,\mathrm{Mpc}}
\newcommand{\mg}{\,\mathrm{mag}}
\newcommand{\magarcsec}{\mathrm{\,mag\,arcsec^{-2}}}
\newcommand{\kms}{\mathrm{\,km\,s^{-1}}}
\newcommand{\s}{\,\mathrm{s}}
\newcommand{\pix}{\,\text{pix}}
\newcommand{\Msun}{M_\odot}
\newcommand{\galfit}{\textsc{galfit}}
\newcommand{\gfthree}{\textsc{galfit3}}
\newcommand{\sex}{\textsc{SExtractor}}
\newcommand{\sersic}{S\'{e}rsic}
\newcommand{\dd}{\mathrm{d}}
\newcommand{\ha}{\mathrm{H}\alpha}
\newcommand{\hi}{\mathrm{HI}}
\renewcommand{\to}{\text{--}}
\def\fsec{\hbox{$.\!\!^{s}$}}

\def\farcsec{\hbox{$.\!\!^{\prime\prime}$}}
\def\farcmin{\hbox{$.\!\!^{\prime}$}}
\def\fdeg{\fdg}

\begin{document}

\title{Discovery of an ultra-diffuse galaxy in the Pisces-Perseus
supercluster\\}

\author{David Mart\'{\i}nez-Delgado\altaffilmark{1,}\altaffilmark{2},
 Ronald L\"asker\altaffilmark{2}, Margarita Sharina\altaffilmark{12},
 Elisa Toloba\altaffilmark{5,}\altaffilmark{13},
 J\"urgen Fliri\altaffilmark{3,15}, Rachael Beaton\altaffilmark{13},
 David Valls-Gabaud\altaffilmark{4}, 
 Igor D. Karachentsev\altaffilmark{12},
 Taylor S. Chonis\altaffilmark{6}, 
 Eva K. Grebel\altaffilmark{1},
 Duncan A. Forbes\altaffilmark{10}, 
 Aaron J. Romanowsky\altaffilmark{5,14},
 J. Gallego-Laborda\altaffilmark{9}, 
 Karel Teuwen\altaffilmark{8},
 M. A. G\'omez-Flechoso\altaffilmark{7},
 Jie Wang\altaffilmark{17},\altaffilmark{11},
Puragra Guhathakurta\altaffilmark{5}, 
Serafim Kaisin\altaffilmark{12},
 Nhung Ho\altaffilmark{16}}

\altaffiltext{1} {Astronomisches Rechen-Institut, Zentrum f\"ur Astronomie der Universit\"at Heidelberg,  M{\"o}nchhofstr. 12-14, 69120 Heidelberg, Germany}
\altaffiltext{2} {Max-Planck-Institut f\"ur Astronomie, K\"onigstuhl 17, 69117 Heidelberg, Germany}
\altaffiltext{3} {Instituto de Astrof\'{\i}sica de Canarias, V\'{\i}a L\'actea s/n, 38200 La Laguna, Tenerife, Spain}
\altaffiltext{4} {LERMA, CNRS UMR 8112, Observatoire de Paris, 61 Avenue de l'Observatoire,  75014 Paris, France}
\altaffiltext{5} {University of California Observatories, 1156 High Street, Santa Cruz, CA 95064, USA}
\altaffiltext{6} {Department of Astronomy, University of Texas at Austin, 2515 Speedway, Stop C1400, Austin, TX 78712, USA}
\altaffiltext{7} {Departamento de Matem\'atica Aplicada (Biomatem\'atica), Universidad Complutense de Madrid, 28040 Madrid, Spain}
\altaffiltext{8} {Remote Observatories Southern Alpes, Verclause, France}
\altaffiltext{9} {Fosca Nit Observatory, Montsec Astronomical Park, Ager, Spain}
\altaffiltext{10} {Center for Astrophysics and Supercomputing, Swinburne University, Hawthorn VIC 3122, Australia}
\altaffiltext{11} {Institute for Computational Cosmology, Department of Physics, Durham University, South Road, Durham DH1 3LE, UK}
\altaffiltext{12} {Special Astrophysical Observatory, Russian Academy of Sciences, Russia}
\altaffiltext{13} {The Observatories of the Carnegie Institutions for Science, 813 Santa Barbara Street, Pasadena, CA 91101 USA}
\altaffiltext{14} {Department of Physics and Astronomy, San Jos\'e State University, One Washington Square, San Jose, CA 95192, USA}
\altaffiltext{15} {Universidad de La Laguna, Departamento de Astrof\'{\i}sica, 38206 La Laguna, Tenerife, Spain}
\altaffiltext{16} {Astronomy Department, Yale University, New Haven, CT06520, USA}
\altaffiltext{17} {National Astronomical Observatories, Chinese Academy of Sciences, Beijing, 100012 China}


\begin{abstract}
We report the discovery of \DGSATI, an ultra-diffuse, quenched galaxy located 10\fdeg 4 degrees in projection from the Andromeda galaxy (M31). This low-surface brightness galaxy ($\mu_V = 24.8\magarcsec$), found with a small amateur telescope, appears unresolved in sub-arcsecond archival Subaru/Suprime-Cam images, and hence has been missed by optical surveys relying on resolved star counts, in spite of its relatively large effective radius ($R_{e}(V) = 12\arcsec$) and proximity (15\arcmin) to the well-known dwarf spheroidal galaxy And II. Its red color ($V-I = 1.0$), shallow S\'ersic index ($n_V=0.68$), and the absence of detectable $\ha$ emission are typical properties of dwarf spheroidal galaxies and suggest that it is mainly composed of old stars. 

Initially interpreted as an interesting case of an isolated dwarf spheroidal galaxy in the local universe,
our radial velocity measurement obtained with the BTA 6-meter telescope ($V_h=5450 \pm 40\kms$)
shows that this system is an M31-background galaxy associated with the filament of the Pisces-Perseus supercluster.
 At the distance of this cluster ($\sim 78\Mpc$),
\DGSATI\ would have an $R_e \sim 4.7 \kpc$ and $M_{V} \sim -16.3$. Its properties resemble those of the ultra-diffuse galaxies recently discovered  in the Coma cluster. \DGSATI\ is the first case of these rare ultra-diffuse galaxies found in this  galaxy cluster. Unlike the ultra-diffuse galaxies associated with the Coma and Virgo clusters, DGSAT I is found in a much lower density environment, which provides a fresh constraint on the formation mechanisms for this intriguing class of galaxy.
\end{abstract}




\section{INTRODUCTION}
\label{sec:intro}


Small aperture ($10 \to 15\cm$) telephoto-lens telescopes, combined with the new generation of commercial CCD cameras, can be valuable instruments to search for low-surface-brightness stellar systems, such as new dwarf companions (Merritt, van Dokkum \& Abraham 2014; Javanmardi al. 2015; Karachentsev et al. 2015; Romanowski et al. 2015) and stellar tidal streams (Mart\'\i nez-Delgado et al. 2008; 2010) around nearby spiral galaxies, and
ultra-diffuse galaxies in galaxy clusters (van Dokkum et al. 2015). The 
short focal ratio of these telescopes allows one to trace
 these faint stellar systems as unresolved, diffuse light 
structures with surface brightness below $\mu_V = 25.5\magarcsec$ well beyond the Local Group, up to distances of $\sim 50 \to 100\Mpc$, exploring the low-surface-brightness regime of the
scaling relations for early-type galaxies. In addition, the single, photographic-film sized chip from such amateur CCD 
cameras coupled with short focal ratio small telescopes can probe extensive sky areas 
with unprecedented depth, reaching surface brightness levels that are $\sim 2 \to 3$ 
magnitudes deeper ($\mu_{r,\mathrm{lim}} \sim 28\magarcsec$) than both the classic photographic plate surveys (e.g., the Palomar Observatory Sky Survey) and the available large-scale digital surveys (e.g. the SDSS). This offers an alternative
and low cost approach for the discovery of low surface brightness galaxies, which may lurk in
those regions where the detection efficiency of the available large-scale imaging surveys drops to very small
values.

Recently, van Dokkum et al. (2015) identified 47 low-surface-brightness galaxies in
the direction of the Coma cluster using the Dragonfly Telephoto Array. Although they have only obtained a  direct
distance for one of these galaxies (van Dokkum et al. 2015b), their spatial distribution suggests that they
are associated with the Coma cluster. Some of these Coma galaxies have a similar size as the Milky Way (which has an effective radius of $3.6\kpc$; Bovy \& Rix 2013), but they are significantly
redder and more diffuse. They are also fainter
than typical low-surface-brightness galaxies ($\sim 24.7\magarcsec$; Bothun et al. 1997), and much
larger than Local Group dwarf galaxies (e.g. the effective radius of
And~XIX, the most extended of the Local Group early-type satellites, is $1.6\kpc$; McConnachie et
al. 2008). These authors dubbed these galaxies
ultra-diffuse galaxies (UDGs) and defined their key properties to include
effective radii greater than $1.5\,\mathrm{kpc}$ and central $g$-band ($V$-band) surface
brightness fainter than $24\,(23.6)\magarcsec$. Although rare, there have been some detections of similar objects in the past. Caldwell et al. (2006) studied two galaxies with similar properties based on their initial
discovery by Impey et al. (1988). Based on the tip of the red giant branch (TRGB) distances, these two galaxies are in the core of the Virgo cluster. Both have an effective radius of $\sim 1.5\kpc$ and a central $V$-band surface brightness of $\sim 26.5\magarcsec$. More recent deep imaging surveys of the Virgo cluster (Ferrarese et al. 2012; Mihos et al. 2015) and the Fornax cluster
(Mu\~noz et al. 2015) have also revealed the presence of faint systems with properties analogous to the UDGs, suggesting that
they could belong to a new morphological class of galaxy with the sizes of the giants but the luminosities of the dwarfs. 

The origin of these ultra-diffuse galaxies is not clear. Whether they are
only found in clusters or whether they also exist in isolation is still 
an open question that --when answered-- will provide important clues about their formation. 
van Dokkum et al. (2015) noted that the spatial distribution of their 47
UDGs avoided the central regions of the Coma cluster, although this apparent location may be an
observational selection effect. The large physical sizes of UDG galaxies compared their low stellar masses
 suggest that they must be dark matter dominated to survive within dense cluster environments.
 If large numbers of UDGs are found in a range of environments, they may contribute meaningfully to the `missing
satellite' problem (e.g., Moore et al. 1999).

During the commissioning of our new project to search for faint satellites around the Local Group spiral galaxies and other nearby systems with amateur telescopes (the \DGSAT\footnote{{\it Dwarf Galaxy Survey with Amateur Telescopes}} survey; see Javanmardi et al. 2015), we found  an apparently isolated, faint  galaxy within the projected extent of the M31 stellar halo, which had been missed by previous surveys that relied on either resolved star counts or the H{\sc i} 21 cm-radio line for galaxy detection. Although it was 
initially interpreted as an interesting case of an isolated dwarf spheroidal galaxy, our follow-up spectroscopic observations confirm that this galaxy is at a distance similar to that of the Pisces-Perseus supercluster of galaxies that is in the background of this field of the M31 stellar halo. The properties that we obtain for this background system are very similar to those of the ultra-diffuse galaxies recently reported by the Dragonfly team (van Dokkum et al. 2015). 

In Section 2 of this paper we describe the discovery of the galaxy and our observing strategy. In Section
3 we detail the photometric and spectroscopic properties that we measure for this galaxy.
In Section 4 we discuss the possible nature of \DGSATI\ as an ultra-diffuse
galaxy and speculate about its possible origin.


\section{OBSERVATIONS AND DATA REDUCTION }
\label{sec:data}


\DGSATI\ was found in a visual inspection of a full color 
image of the Andromeda~II dwarf galaxy available on the internet and 
taken by the amateur astronomer
 Alessandro Maggi using a Takahashi Epsilon 180ED astrograph ($18\cm$ 
diameter at $f/2.8$).  The detection was confirmed by follow-up 
observations using the amateur and professional facilities described 
below. The position of the center of this new dwarf galaxy lies at 
$\alpha_0 = 01^h \,17^m \, 35\fsec59$ and  $\delta_0 = +33^\circ \, 31\arcmin \, 42\farcsec37$ (J2000).


\subsection{Amateur telescope datasets}
\label{subsec:amateur}

The first dataset was collected with a 15-cm aperture $f/7.3$ Takahashi TOA-150 refractor telescope located at the Fosca Nit 
Observatory (FNO) at the Montsec Astronomical Park (Ager, Spain). We used a STL-11000M CCD camera with a large FOV ($1\text{\fdeg}9 \times 1\text{\fdeg} 3$, $1 \farcsec 69\pix^{-1}$). A set of individual $1200\s$ (luminance-filter) 
images were obtained in photometric nights between August and September 2012, with a total integration of 43,800 seconds.
Each sub-exposure was reduced following standard procedures for dark subtraction, bias correction, and flat fielding.

Deeper imaging of the field was acquired with a 0.4-m aperture, $f/3.75$ 
corrected Newton telescope located at the Remote Observatory 
Southern Alps (ROSA, Verclause, France) in August 2013. We used a FLI ML16803 CCD detector, which
 provides a total field of $81\arcmin \times 81\arcmin$ with
 a scale of $1\farcsec 237\pix^{-1}$. The total exposure 
time in the  luminance filter was $13,200\s$ and the images
 were reduced using the same standard procedure. 

\begin{figure*}[!htb]
\centering{\includegraphics[scale=0.8, angle=0]{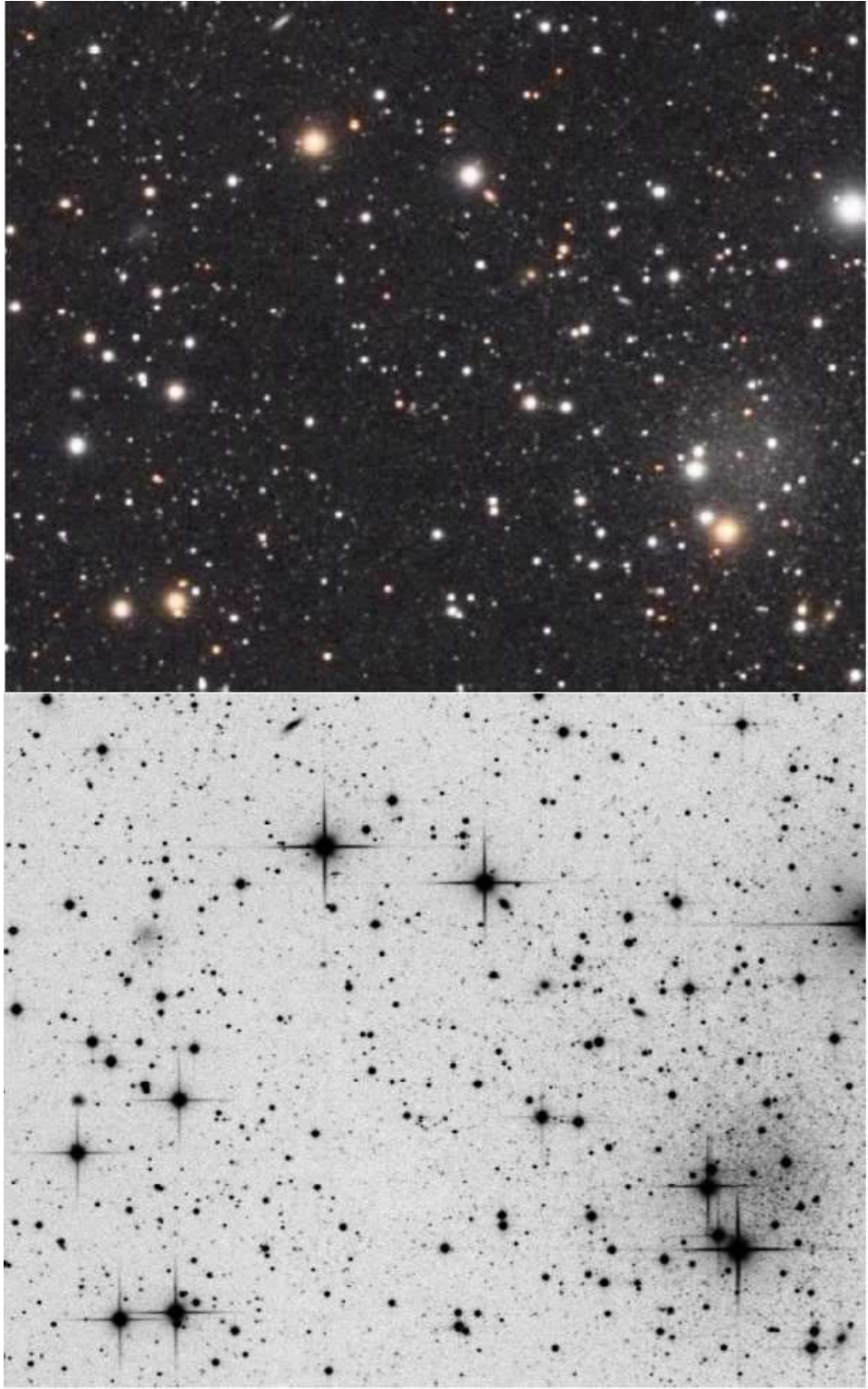}}
\figcaption{Follow-up small telescope images of \DGSATI : ({\it top}) color image obtained with
the FNO TOA-150 refractor; ({\it bottom}) luminance filter image  
obtained with the ROSA 0.4-meter telescope. The new dwarf is detected
as a small cloud (top left) close to the And II dSph (bottom right),
 visible only $\sim$ 15\arcmin~ to the West. North is top, East is left. 
The field of view of these cropped images is $\sim$ 19\arcmin $\times$ 
11\arcmin. \label{fig1}}
\end{figure*}

Both images obtained to confirm the discovery of \DGSATI\  are shown in Figure~\ref{fig1}. The luminance images taken with these amateur telescopes were first 
astrometrically calibrated \citep{lang10}, and then calibrated to the SDSS-DR7 \citep{2009ApJS..182..543A} photometric system. To begin 
the flux calibration of both images, we remove any residual large-scale 
sky gradients in our wide-field images by modeling the background using a two-dimensional, fourth-order Legendre polynomial that was fitted to 
the median flux in coarse pixel bins after masking all sources that were 
detected above the background at $\gtrsim 5\sigma$ level. The bins were 
typically 5\arcmin\ on a side. 

We then located 67 isolated stars distributed throughout each image 
with SDSS $r$-band magnitude $16<r<19$, and performed aperture photometry on them. The same 
set of stars was measured in both images. The positions of these stars 
were matched with the SDSS DR7 catalog, from which we retrieved their $g$, 
$r$ and $i$ magnitudes. From these data, a linear relation between 
the $r$ magnitude and the luminance
instrumental magnitude was derived individually for each image. We 
find that the residuals from this relation are a strong linear function 
of $g-i$ color and vary by 0.5 mag over the color range 
$0.4 < g-i < 2.7$. Once this color term was corrected, 
the statistical uncertainty in the flux calibration is only $0.04\mg$. 
The $5\sigma$ limiting surface brightness for background variations measured following
the procedure of Cooper et al. (2011; see their appendix)
is $27.16\magarcsec$ and $26.03\magarcsec$ for the 
TOA-150 and 0.4-m $f/3$ data, respectively, as determined by measuring 
the standard deviation of the median of random 20\arcsec\ apertures placed
around each image. This means that the TOA-150 
imaging is $\sim1\mg$ deeper than the ROSA 0.4-m data in terms of 
photon statistics, and its sensitivity to large-scale surface brightness
 variations is less hindered due to the better quality of the flatfielding. 


\subsection{Subaru optical archive data}
\label{subsec:subaru}

We use a deep image of the field around the And~II dSph obtained  with the  Subaru/SuprimeCam wide field imager \citep[$34\arcmin \times 27\arcmin$ FOV, $0\farcsec202\pix$;][]{2002PASJ...54..833M}, which is publicly available from 
the SMOKA archive \citep{2002RNAOJ...6...23B}. These observations include dithered exposures of $5\times440\s$ in the Subaru $V$-band and $20\times240\s$ in the Subaru $I$-band with a seeing around 0\farcsec6 \citep{2007MNRAS.379..379M,2012ApJ...758..124H}. 

\begin{figure*}[!htb]
 \centering{\includegraphics[scale=0.80]{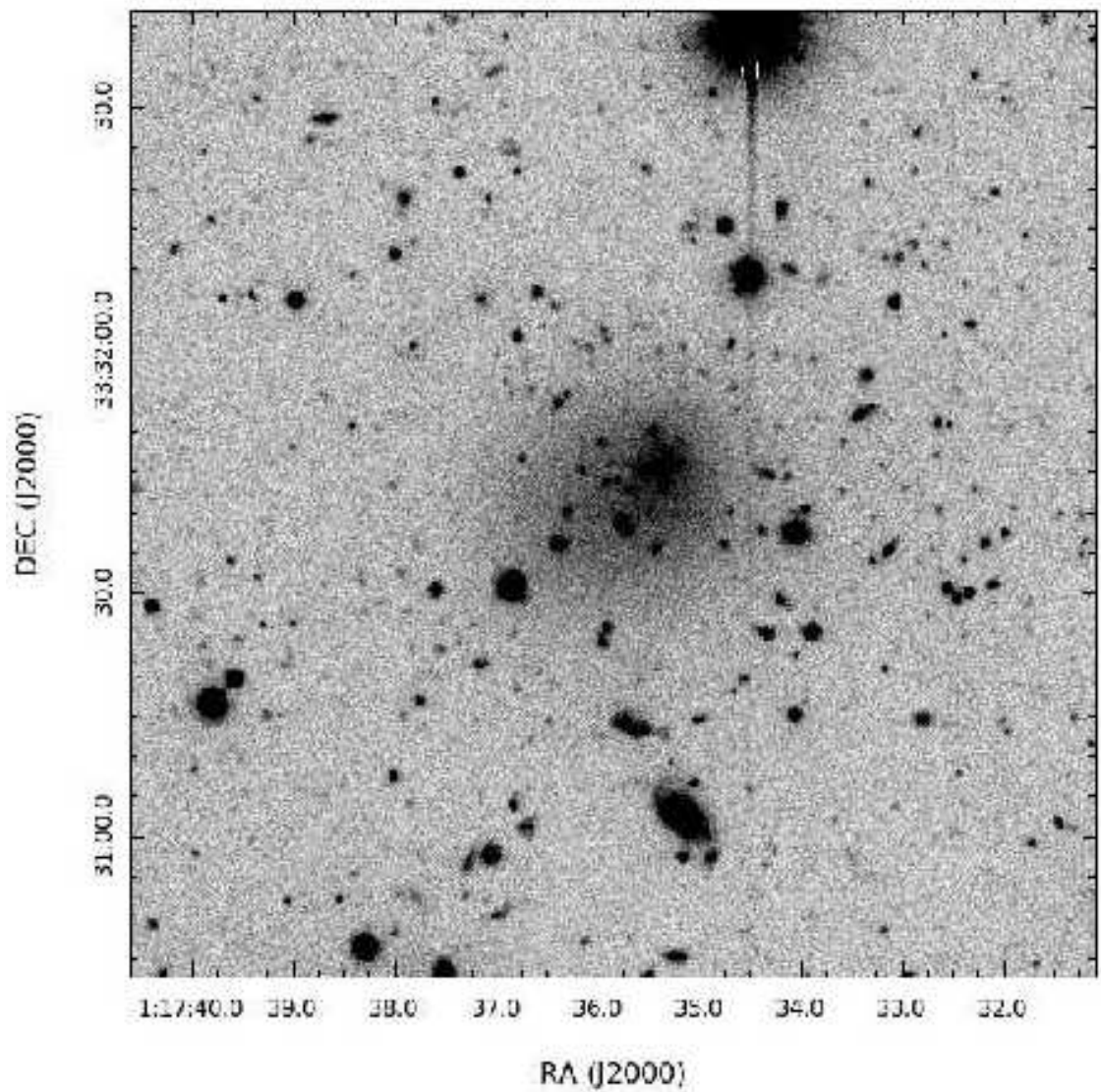}}
\figcaption{Subaru/SuprimeCam $V$-band image of \DGSATI\
 from the SMOKA archive (see Sec.~\ref{subsec:subaru}). \label{fig2}}
\end{figure*}

Preprocessing of the data was done by debiasing, trimming, flat fielding, and gain correcting 
each individual exposure chip-by-chip using median stacks of nightly sky flats.
The presence of scattered light due to bright stars both in and out of 
the field of view required removing this smoothly varying component before performing photometry 
and solving for a World Coordinate System solution. To remove scattered light, we fitted 
the smoothly varying component by creating a flat for every chip within each frame by 
performing a running median with a box size of 300 pixels. This was then subtracted from the original,
unsmoothed frame to produce a final image for photometric processing. The resulting Subaru $V$-band image is given in Figure~\ref{fig2}.

Photometric calibration was performed using the \textsc{daophot~ii} and \textsc{allstar} 
\citep{1987PASP...99..191S} packages. Sources were detected in the $V$-band image by requiring a $3\sigma$-excess above
the local background, with the list of detections being 
subsequently used for the \textsc{allstar} run on the $I$-band image. To obtain clean CMDs and minimize the contamination by faint background galaxies, sources with the ratio estimator of the pixel-to-pixel scatter $\chi<2.0$ 
and sharpness parameters $|S| < 1.0$ \citep[see][]{1987PASP...99..191S} were kept for 
further analysis. Instrumental magnitudes were obtained by cross-correlating with
stellar photometry in the Subaru filter system from McConnachie et al. (2007), kindly provided in digital form, with zero-points
determined as resistant mean of the differences between our instrumental and the calibrated Subaru magnitudes (denoted here as V',I'). The accuracy (standard error) of the zero-point is better than $0.03\mg$. For the transformation to the standard system we combined equation 1 in McConnachie et
al. (2007), which link the Subaru and the INT filter systems, with the
transformation equations for the INT and Johnson/Kron-Cousins systems,
available on the INT Wide Field Survey (WFS) webpage\footnote{http://www.ast.cam.ac.uk/~wfcsur/technical/photom/colours}.
The resulting transformations relations are:
\begin{eqnarray*}
V &=& V'+0.040\times(V'-I') \\
I &=& I'-0.090\times(V'-I')~,
\end{eqnarray*}
where ($V$,~$I$) and ($V'$,~$I'$) are the magnitudes in the standard and Subaru
systems, respectively.


\subsection{Spectroscopic observations}
\label{subsec:spec}

During the refereeing process of this paper, spectroscopic observations of \DGSATI\ were obtained using the primary focus of the
\mbox{6-m} Bolshoi Teleskop Alt-azimutalnyi (BTA) telescope of the Special Astrophysical Observatory of the 
Russian Academy of Sciences (SAO RAS)
with the SCORPIO spectrograph\footnote{http://www.sao.ru/hq/lon/SCORPIO/scorpio.html} \citep{2005AstL...31..194A} on October 28th-29th, 2014. The slit width was $1\arcsec$ and its length was $6\arcmin$. 
The instrumental setup included the CCD detector EEV\,42-40 with a pixel scale of 0.18\,arcmin\,pixel$^{-1} $
and the grism VPHG1200B (1200\,lines\,mm$^{-1}$) with a resolution of ${\rm FWHM}\sim 5$\,\AA, a reciprocal dispersion of $\sim 0.9$\,\AA\ per pixel and a spectral range of 3700--5500\,\AA. Two spectra of 12000 and $13200\s$ exposure time were obtained in two consecutive nights under very good atmospheric conditions. At the beginning and at the end of each night a sequence of 10--20 bias frames was recorded. The fluctuations of the bias level were small ($\sim 4 {\rm e}^-$).  
Flatfield and He-Ne-Ar arc calibrations were taken each science exposure, which allowed us to provide a reliable wavelength calibration and a pixel-by-pixel sensitivity correction for our spectra. 

The data reduction was performed using the European Southern Observatory Munich Image Data Analysis System \citep[\textsc{MIDAS;}][]{1983Msngr..31...26B} and the Image Reduction and Analysis Facility ({\sc IRAF}) software system (Tody 1993). It included bias and dark subtraction, flat-field correction, wavelength calibration,
background subtraction and extraction of one-dimensional spectra. 
The dispersion solution provided an accuracy of the wavelength
calibration of $\sim 0.16$\,\AA.  
Because our object is of very low surface brightness, accurate subtraction of night sky lines is critical for the correct determination of its radial velocity.
The sky background subtraction and correction of the spectrum curvature were done using the {\it background} task in {\sc IRAF}. Finally, one-dimensional spectra were extracted from the two-dimensional ones and summed. The signal-to-noise ratio in the integrated light spectrum of \DGSATI\ is $\sim 5$ per wavelength bin and per spatial resolution element. The resulting integrated spectrum is displayed in Figure~3. 


\subsection{SAO narrow-band observations}
\label{subsec:ha}

Narrow-band observations were made with the 6-m telescope at the SAO RAS
using the SCORPIO detector with a $2048 \times 2048$ pixel matrix in 
a $2 \times 2$ binning mode and an image scale of 0.18\arcsec/pixel, which yields a full 
field of view of 6\farcmin1$\times$6\farcmin1. Images in $\ha$+[N\textsc{ii}] and in 
the continuum were obtained on October 26, 2013 by observing 
the galaxies through a narrow band $\ha$ filter (${\rm FWHM}=75$\,\AA) with an effective wavelength 
of 6555\,\AA, and the SED607 (with ${\rm FWHM}=167$\,\AA, $\lambda=6063$\,\AA) and 
SED707 (with ${\rm FWHM} = 207$\,\AA, $\lambda = 7036$\,\AA) intermediate band filters for the 
continuum. The exposure times were $6 \times 300\s$ in the continuum and 
$3 \times 600\s$ in $\ha$. 

A standard procedure for analysis of direct CCD 
images was used for processing the data. The bias was initially
subtracted from all the data, and then all the images were divided 
by a flat field. Afterwards, the cosmic ray events were removed and the 
sky background was measured and subtracted from each image. The continuum images 
for each pointing were normalized to the corresponding $\ha$ images using 15 field
stars so that continuum could be subtracted from the $\ha$ images. The $\ha$  flux 
was then measured from the continuum 
subtracted images and calibrated using spectrophotometric exposures of standard 
stars obtained on the same night. For \DGSATI\ we detected no $\ha$ 
emission, with an upper limit for its $\ha$ flux of 
$\log F(\ha)\,[{\rm erg}\s^{-1}\cm^{-2}] < -15.8$. This limit will be
used to set constraints on the stellar populations in Section~\ref{subsec:mass+sfr}.


\section{RESULTS}


\subsection{Color-Magnitude Diagrams}
\label{subsec:cmd}

Figure~\ref{fig4} shows the CMD of a selected region of 
$8\farcmin15 \times 15\farcmin03$ from the
Subaru field (corrected for Galactic extinction A$_{B}$=0.27) 
with RGB stars 
from the halo of M31 and some Galactic foreground
stars at $(V-I)_0 \ge 2.0$. Based on the
densely populated CMDs from the PAndAS survey 
\citep{2009Natur.461...66M} of M31's halo \citep[e.g., Figure~4 in][]{2011ApJ...740...69C}, 
we selected stars with $(V-I)_0 \le 2.0$
for our sample, eliminating most of the Galactic
foreground population but still keeping most of the stars in M31's halo 
and potential resolved
stellar sources in the dwarf galaxy. Using a similar color criterion as 
Conn et al. (2012) to select putative RGB
stars in the satellite candidate (Figure~\ref{fig4}), we find that 
only 17 stars fall in this
area of the CMD and within 40\arcsec\ of the center of the galaxy, 
consistent with the
average density of stars within these color cuts in the image. Hence, 
no significant over-density
of resolved RGB stars is found at the location of \DGSATI\ that would indicate it is
a dwarf satellite of M31. This
is consistent with the lack of detection of this object in the list of 
over-density candidates in this region of Andromeda from the PAndAs survey 
\citep{2013ApJ...776...80M}. Thus, we conclude that the galaxy is not resolved into stars. This further rejects the possibility that \DGSATI\ is a companion of the more distant spiral galaxy  NGC~404 that is 2\fdeg75 distant in projection \citep[$\sim 3.13$ Mpc;][]{Williams2010}. It is also very unlikely that it is a dwarf member of any other galaxy group in the background of M31 in the local universe (e.g. NGC~672 with 5 known  companions or NGC~891 with 18 known companions). 


\subsection{Radial velocity and Distance}
\label{subsec:vel+dist}

To derive radial velocities of our object we used: a) the {\it fxcor} task in {\sc IRAF}; and b) the {\sc ULySS} program
\citep{2008MNRAS.385.1998K, 2009AA...501.1269K} with Vazdekis et al.'s (2010) SSP model, the Salpeter IMF (Salpeter 1955) and the MILES stellar library \citep{2006MNRAS.371..703S}.
The task {\it fxcor} uses the Fourier cross-correlation method developed by Tonry \& Davis (1979). The resulting radial velocity
is $V_h=5453 \pm 111\kms$. The cross-correlation peak is weak and the velocity error is large because of the low $S/N$ in the spectrum.

The {\sc ULySS} program provides more accurate results than {\it fxcor}, because it allows one to take into account
the line-spread function (LSF) of the spectrograph, viz.
the variation of the measured velocity and instrumental velocity dispersion as a function of wavelength (Koleva et al. 2008). 
The LSF was approximated by comparing a model spectrum of
the Sun to twilight spectra taken during the same night as the studied object.
The result is $V_h=5450 \pm 40\kms$. Using the prescriptions of Tully et al. (2008), we transform our initial measured velocity to
one relative to the motion of the Local Group and obtain $V_{LG}=5718 \pm 40\kms$. This radial velocity corresponds to the Hubble distance of $78 \pm 1\Mpc$ for $H_0=73\kms\Mpc^{-1}$. The estimated velocity dispersion in the galaxy is $\sigma_V \sim 100\kms$.
 The $\sigma_V$ value is very approximate given the low $S/N$ in the spectrum of \DGSATI\ and needs to be verified at higher resolution and at higher S/N.

Interestingly, this velocity is consistent with those of
the Pisces-Perseus supercluster, which has typical velocities in the range of $4800 \to 5400\kms$. Figure~5 shows the position of \DGSATI, clearly overlapping a prominent filament of this supercluster in this sky region
(Wegner, Haynes \& Giovanelli 1993) and in proximity to some of the massive members of this structure (NGC 383 and NGC 507, at projected distances of 2-3 Mpc). Therefore \DGSATI\ could well be part of this structure, lending support to our redshift distance estimate, with a corresponding diameter (effective size) near $\sim 10\kpc$ (see next section). Its simultaneous low surface brightness ($\sim 25\magarcsec$) and large physical extent would place it in the category of the {\it ultra-diffuse} galaxies (see Sec. 4).


\subsection{Structural properties and surface brightness profile}
\label{subsec:galfit}

The structural parameters of the dwarf galaxy were determined by
running \gfthree\ \citep{2010AJ....139.2097P} on the sky-subtracted Subaru $V$- and $I$-band images, 
with foreground and background objects removed. We characterize the 
photometric
properties of the galaxy by using \gfthree\ to fit a two-dimensional 
\sersic\ profile,
\begin{eqnarray*}
       I(R) &=& I_c \exp\left[-b_n\left(R/R_e\right)^{1/n}\right] \\
 I_c = I(0) &=& F_\mathrm{tot} \; \left[2\pi 
q n \left(R_e b_n^{-n}\right)^2 \Gamma(2n) \right]^{-1} \; ,    \label{eqn:sersic}      
\end{eqnarray*}
\noindent where $I$ is the surface brightness within ellipses with 
semi-major axes $R$, and $\Gamma$ is the Gamma function. The conversion 
to the magnitude 
system (see Table~1) is given by 
$\mu(R)=\mathrm{ZP}-2.5\log I(R)$, 
where $\mathrm{ZP}$ is the magnitude zero-point and $\mu_c=\mu(0)$ the 
central surface 
brightness. The total flux $F_\mathrm{tot}$ (expressed as magnitude, $m=\mathrm{ZP}-2.5\log F_\mathrm{tot}$), the effective radius $R_e$, \sersic\ index $n$, and axis ratio $q=b/a$ 
are adjustable parameters. The constant $b_n$ is determined by the requirement 
$F(<R_e) \propto \int_0^{R_e} I(R)\,R\,\dd R = F_\mathrm{tot}/2$. 
The geometry of the light distribution is additionally characterized by the location of the profile center and the position angle of the semi-major axis.

 We perform masking of intervening objects (mostly foreground
stars) and background subtraction iteratively with \sex\ \citep{sextractor} in the area where we fit our model ($280'' \times 280''$), and add user-defined masked regions around the brightest stars. In order to obtain the background reliably, we also mask the target galaxy generously "by hand". After a first pass of simultaneous object detection and background estimation, the task is repeated with the previously detected objects masked, which gives us a more accurate local background level and background noise, and thus also a more complete object mask. Finally, by reapplying \sex\ on the \galfit\ residual image, we eventually also mask previously undetected objects that overlap with our target.

The noise map, which enters the $\chi^2$ computation during fitting, was  constructed by measuring the noise in the background and adding in quadrature the Poisson noise of the sources. In lieu of including the sky background as a free parameter in the \galfit\ model, which can induce degeneracy in the output model parameters, we subtract the background from the image before fitting. Lastly, to account for the effect of seeing and telescope optics, we construct a model image of the PSF using several bright, unsaturated foreground stars near the position of \DGSATI\ in the image. This empirically derived PSF model is convolved with the \galfit\ model before comparison to the image data. 

We first fit a 2D-\sersic\ profile as a basic model of the $V$- and $I$-light distribution of \DGSATI\.In this simple model, we find \DGSATI\ is almost circular ($b/a \sim 0.9$) with a shallow, sub-exponential inner profile ($n<1$), both of which are typical structural characteristics for galaxies of dwarf spheroidal morphology. However, despite the overall agreement of the model with the data (see Figure~6), the data show two peculiarities with respect to the smooth regular \sersic\ model.

First, there is a faint, fuzzy irregularly-shaped elongated over-density of
$\sim 6''$ length near, but not coinciding with, the 
galaxy center (offset by $\sim 1''$). Apart from its shape, the over-density also differs from the main body of the galaxy by its bluer color, more specifically $V-I=0.7$ for the over-density as compared to $V-I=1.0$ for a regions at a similar radius within \DGSATI\. Due to its small angular size and the spatial resolution of our ground-based imaging, it is not possible to shed additional light on the nature of the blue over-density at this time (see also Sec.~3.4) . It could be an unresolved clump of young stars  overlapping the smooth (and possibly older) component of the galaxy, similar to those observed in some nearby dIrr galaxies (e.g. Sextans A, that contains a conspicuous, off-center star formation region; Dohm-Palmer et al. 1997). Alternatively, DGSAT~I could be a nucleated UDGs, similar to those found in the Virgo cluster (Mihos et al. 2015; Beasley et al. 2015).

 For this reason, we construct three \gfthree\ models that differ in how 
this blue off-center over-density is treated. Model 1 assumes that it should  be included in the fit, and leaves it unmasked. 
Model 2 also leaves it unmasked, but accounts for it as an additional structural component 
with fully independent parameters from the main body of the galaxy (i.e., both the general parameters and those specific to the \sersic\c fit). In Model 3 we exclude the over-density from the fit by masking it. Table~1 compares  the 
resulting parameters of the main component (the sole component in Model 1 
and Model 3). Parameters generally vary only mildly between these models, by $\sim 0.01$ 
for $m$, $R_e$ and $q=b/a$. Differences in the best-fit $\mu_c$ and $n$ are more 
pronounced ($\sim 0.1$) as they more specifically reflect the central profile.

Another peculiarity in the light distribution of \DGSAT\ is its significant lopsidedness: the center of the innermost isophotes are offset from those of the outermost isophotes by several arcseconds to the north-west. We accommodate for the apparent lopsidedness by allowing a first-order Fourier mode in the model isophotes. The best-fit amplitude of the mode is 0.21 (0.15) in $V$-band ($I$-band). Although allowing for the Fourier mode in a single-component system is not essential to obtain accurate flux and scale parameters, this modification to the model is useful for us since it provides a measurement of the galaxy center and a realistic set of isophotes, which we utilize to measure the surface brightness profile $\mu(R)$ with the IRAF task \textit{ellipse} (see below). 

When measuring $\mu(R)$ on the data image, we use the isophotes of the model. Fitting the isophote geometry simultaneously with $\mu(R)$ (which was done on the model image) is impeded at small radii due to the low central surface brightness, the extremely shallow central gradient, and the irregularities introduced by the off-center over-density and remaining small contaminants. 

For the following discussion of the DGSAT~I properties, we select Model~3 (where the over-density is masked) as our adopted model. We do not base this selection on the value of $\chi^2$ at the best-fit parameters, since it is very similar for all three models considered\footnote{Regardless, the minimum-$\chi^2$ of different models generally does not provide statistically well-defined quantitative evidence for intercomparison of models}. We assume that, given its different color and presumably younger population, this over-density is a peculiar feature of DGSAT~I and  its light is not reflective of the global properties (including the stellar mass) at  same 
confidence as as in the smooth, red component of the galaxy. 
A {\it rough} estimate of the flux and size of this over-density can be obtained from the extra-component fitted in Model 2, yielding  2.5\% flux fraction in the $V$-band (1.1\% in the $I$-band), and 22\% of the effective radius (both bands; see Table 2). 

The apparent galaxy parameters (Table 1, column for "Model 3") translate to physical parameters using our adopted redshift distance of $78\Mpc$. The extinction-corrected apparent $I$-band magnitude of $17.17$ implies an absolute magnitude, $M_I=-17.29$, and, adopting $M_{\odot,I}$=4.1 (Binney \& Merrifield 1998), a luminosity $L_I=3.6 \times 10^8 L_{\odot,I}$. In the $V$-band, we obtain $m_V=18.18$, which implies an absolute magnitude, $M_V=-16.28$ and, with $M_{\odot,V}=4.8$, $L_V=2.7 \times 10^8\,L_{\odot,V}$. At the same time, the effective radius of $12\farcs5$ translates to $4.7\kpc$. There is some uncertainty in $R_e$, both due to the variation with band as well as due to background uncertainty; we give here the maximum range of $R_e$ resulting from fitting different models {\it and} both bands (see above and Table 1), which is $(13 \farcs 1-11 \farcs 3)/2=0 \farcs 9$, hence $\approx 10\%$ or $\approx 0.5\kpc$ maximum range in $R_e$. Thus, this galaxy is very extended (about Milky-Way sized; Bovy \& Rix 2013) for its luminosity (which is less than 1/100 of the Milky Way). 


\subsection{Stellar Mass and Star Formation Rate}
\label{subsec:mass+sfr}

In order to estimate the stellar mass of \DGSATI\, we use the work of Zibetti, Charlot \& Rix (2009), who provide mass-to-light ratios ($M_\star/L$) in SDSS bands for galaxy stellar populations as a function of one or more colors. They also show that the mass-to-light ratio in the $i$-band is well
determined (low scatter and low dust effects) by one color alone, $g-i$. Fortunately, our Subaru images provide photometry in closely related bands, however we still need to transform the Johnson-Cousins system to the SDSS system.

With only one color at hand, we have little knowledge of the stellar population parameters (primarily age and metallicity) that effect the transformation between photometric bands. We refer to Jordi, Grebel \& Ammon (2006) when calculating $g-i = 1.481 \times (V-I) - 0.536 = 0.960$ with our measured $V-I=1.01$; this transformation is based on Stetson's extension of the Landolt standard stars and hence a mix of stellar populations. A transformation from Johnson $I$ to SDSS $i$ is unfortunately not possible from the tables provided in Jordi et al.(2006); we rely here on the relations of Lupton (2005), which are available on the SDSS website\footnote{https://www.sdss3.org/dr8/algorithms/sdssUBVRITransform.php} and use the same set of standard stars. We obtain $m_i \equiv i = I + 0.127 \times (V-I) + 0.320 = 17.62$, corresponding to $\log (L_i/L_{\odot,i})=8.57$ when adopting an absolute solar magnitude of $M_{\odot,i}=4.58$ (Blanton et al. 2003). With $\log \left[ (M_\star/L_i)~/~(M_\odot/L_{\odot,i})\right] = −0.963 + 1.032 \times (g-i)$, we obtain a stellar mass-to-light ratio $M_\star/L_i = 1.07\,M_\odot/L_{\odot,i}$ and a stellar mass for \DGSATI\ of $M_\star = 4.0 \times 10^8 M_\odot$.

We mention that the stellar $M/L$ is an uncertain quantity; amongst other factors it is sensitive to priors in the star formation history and assumptions on the shape of the stellar initial mass function. For example, the relations of Bell et al. (2003) yield $M_\star/L_i = 2.2\,\Msun/L_{\odot,i}$ for the same $g-i$ color; a factor of two higher than the result based on Zibetti et al. For comparison, using either only Population-I stars or Population-II stars, $g-i$ is 0.97 or 0.93, i.e. the stellar population effects the color transformation by $\sim 0.04\mg$, and thus $M/L_i$ by a factor of $\lesssim 0.04$. The $i$-band magnitude is even less sensitive to stellar population differences than the $g-i$ color. For example, using the SDSS-Johnson transformation derived for Population-I stars from Jordi et al., in conjunction with $R-I$ colors from Caldwell et al. (1993), $m_i$ differs by a mere $0.002\mg$ from our adopted transformation using all standard stars.

However, the relatively large systematic uncertainty in $M_\star/L$  does not effect the conclusions of our study: that \DGSATI\ is extremely extended for its mass (or luminosity) and therefore may represent a possible class of peculiar galaxies with an extremely low surface brightness (that is typical of the much smaller dSph galaxies) for their mass (or size) . We therefore attempt to explore further properties of DGSAT-I from the available data, namely $\ha$ flux, star formation rate, and $\hi$ content, that may help to shed light on its formation.

The lack of spatially extended emission lines in the SAO spectra (see Sec.~\ref{subsec:spec}) is consistent with our $\ha$-band observations and rules out the possibility of this object being a nearby star-forming galaxy. This is also consistent with the observed upper limit of $\ha$ flux detected in our SAO narrow band observations (Sec.~\ref{subsec:ha}). We also note that \DGSATI\ was not detected in the 21 cm survey of this filament of the Pisces-Perseus superclusters by Giovanelli \& Haynes (1989). Taking the upper limit for the flux of this survey, $F(\hi)=0.5\,{\rm Jy}\kms$ (with signal-to-noise $S/N \sim 5$), this yields $\log M_{HI}[M_\odot]< 8.76$\footnote{\DGSATI\ is also within the ALFALFA survey zone,
but the HI data for this declination are not yet available.}. We hence estimate the logarithm of the ratio of $\hi$-to-total stellar masses, $\log M_\hi / M_\star < 0.16$. Accepting this upper limit,  implies about as much mass in $\hi$ as in stars. This is, for example, four times $\hi$ fraction than that found in the LMC (Kim et al. 1998), which has one of the highest $\hi$ fractions of the galaxies in the local universe. Therefore we conclude that the upper-limit estimate is a very conservative one, and presume that $M_\hi$ of \DGSATI\ is actually at least one magnitude lower.

Following \citet{Kennicutt1998}, we determined the integral star-formation rate (SFR) in \DGSATI\ by the relation,

\begin{equation}
\log \mathrm{SFR} [M_\odot\,{\rm yr}^{-1}] \; = \; 
8.98 + 2 \log(D[{\rm Mpc}]) + \log(F_{c}(\ha))
\end{equation}

\noindent
where $F_{c}(\ha)$ is its integral flux $F$ in the $\ha$ line in $\mathrm{erg\,cm^{-2}\,s}$ (see Section \ref{subsec:ha}), corrected for the Galactic extinction $A(\ha)$ = 0.538 $A_{B}$ \citep[][with $A_B$=0.27]{schlegel98}. The internal extinction in the dwarf galaxy itself was considered negligible. Therefore, for the distance of \DGSATI, we obtain an upper limit of $\log(\mathrm{SFR}[\Msun\,{\rm yr}^{-1}]) < -2.56$. The corresponding upper limit for the specific star formation rate (sSFR; the SFR per unit galaxy stellar mass) is $\mathrm{sSFR} < -11.16$.

Karachentsev \& Kaisina (2013) discussed star formation properties of Local Volume galaxies. The SFR of \DGSATI\ is somewhat low for its upper limit $\hi$ mass, but still consistent with the spread in their sample (see their Figure~5). They showed that the median
value of the sSFR does not change significantly for local volume galaxies with neutral hydrogen masses in the range $7<\log M_\hi [M_\sun] <9.5$. Most galaxies have $\log\mathrm{sSFR}[{\rm yr}^{-1}] \sim -10$, but with a large scatter towards the lower sSFR over the whole range of $M_\hi$. Some objects with neutral hydrogen masses in the range $7<\log M_\hi [M_\sun] <9.5$ reach logarithmic sSFRs as low as $\sim -15$. Thus, the upper-limit sSFR and SFR/  $M_{HI}$ ratio of \DGSATI\ is consistent with their Local Volume study, and comparable with the values typical for quenched galaxies. However, SFR or hydrogen mass may also be much lower, a possibility that will have to await additional data to be tested.

\DGSATI\ displays a clear off-center over-density discussed in Sec.~\ref{subsec:galfit} (clearly visible in Figure~2 and Figure~7). This raises the question of whether this feature harbors a young stellar population, originating in its last episode 
of star formation. Interestingly, irregardless of the our treatment of the blue over-density in our \gfthree modelling (Table 1), the central $V-I$ color of the \sersic\ profile is still $\sim0.2\mg$ bluer than its global mean color (compare the $V-I$ rows in the top two sections, $m$ and $\mu_{c}$,  in Table~1). It is not clear if this difference is real or an artifact from profile mismatch or background uncertainties. Thus, we perform an additional test by comparing the mean color in two different regions: (1) a circular annulus between  $4''<R<15''$ from the center of \DGSATI\; and (2) a small elliptical aperture at the center with semi-major and -minor axes $a$=1\farcsec4, $b$=0\farcsec6 with a position angle of $PA=10\deg$ North-West. These two annuli are indicated in the left panel of  Figure~7(and described in the caption). As can be seen in Figure~7, these two regions are largely selected based on the features of the $V-I$ color map, with the region (1) selected to both average over background fluctuations but also to avoid contamination from the blue
over-density and region (2) selected to sample the core of the object. From this analysis, we find for region (1) $V-I=1.00$ and for
region (2) $V-I=0.71$, which broadly confirms the bluer nature of the center as measured in the \gfthree\ modelling (Table 1).
To ascertain the significance of this finding and evaluate any subjectivity in the definition of the regions, we perform a Monte-Carlo re-analysis with $N=400$ samples, where the (i) center and (ii) specific parameters for the region shapes are allowed to vary. More
specifically, for region (1), the inner and outer radius and for region (2), 
 the semi-major and semi-minor axes ($a$, $b$) as well as the position angle. 
The center is allowed to shift by 20\% of the harmonic axis mean 
(0.2$\sqrt{ab}$), the direction of the shift is random, $R_{in}$, 
$R_{out}$, $a$ and $b$ are varied by $\pm20\%$ of their given value, 
and $PA$ by $\pm10\deg$. All variations are independent of one another and uniformly 
distributed within their bounds. As a result, the geometric color 
uncertainty is small, with a standard deviation of $0.01(0.02)\mg$ 
in the outer (central) aperture, which confirms that the bluer color for the center is robust.

It is difficult to assess with the current data what this 
$0.3\mg$ color difference in the center implies about a possible stellar population gradient. The color offset, location and shape of the over-density is similar to those reported for dwarf elliptical (dE) galaxies by Lisker et al. (2006), which were interpreted as the presence of young stars overlaying the mass-dominant old population. It is therefore likely that the bluer color of the over-density feature of \DGSATI\ ($V-I \sim 0.7$) indicates the presence of young stars from a recent episode of star formation in this region of the galaxy. 


\section{DISCUSSION AND CONCLUSIONS}
\label{sec:disc+concl}

We report the discovery of \DGSATI\,  a faint galaxy at a projected distance of 
15\arcmin\ from the Andromeda~II dwarf satellite. Its appearance, structural
properties and absence of emission lines initially suggested an interesting case of an isolated dwarf galaxy well beyond the Local Group, with a surface brightness and structural properties similar to those of the classical MW dSphs like Fornax or Sculptor. However, our measured line-of-sight radial velocity ($5450 \pm 40\kms$) reveals that \DGSATI\ is a background system placed at a distance of $\sim 78\Mpc$ and possibly associated with the filament of the Pisces-Perseus supercluster projected in this direction of the sky (see Figure~\ref{fig:image}).

With $\mu_{c,V}=24.8\magarcsec$, $R_e=4.7\kpc$, $b/a=0.9$ and $V-I=1.0$, the central surface brightness, structural properties, and color are consistent with those of the ultra-diffuse galaxies recently reported in the Coma cluster
(van Dokkum et al. 2015; Koda et al. 2015). These galaxies have $g$-band central magnitudes of $24 \to 26\magarcsec$, effective radii of $1.5 \to 4.5\kpc$ and appear nearly round on the sky with typical axis ratios of 0.8, and $g-i \sim 0.8$. van Dokumm et al. (2015) fitted a a S\'ersic index of $n=1$ to their surface brightness profiles, but noted that the data were fit equally well with indices varying from n=0.5 to n=1.5. In the case of \DGSATI, our Subaru images are deep enough for a robust 2D-profile fit, which yields $n \sim 0.6$. The absence of a clear disk structure or spiral arms in the Subaru deep image of \DGSATI\
 supports van Dokkum's suggestion that these objects do not resemble the classical low-surface brightness galaxies well known in the literature.  This is also consistent with the gas content and integral SFR of \DGSATI\ that we found in Sec.~\ref{subsec:mass+sfr}, which are typical for quenched galaxies.

 Thus, \DGSATI\ is the first ultra-diffuse galaxy found in the Pisces-Perseus supercluster. However, unlike these previous detected UDGs, DGSAT~I is almost a {\it field} UDGs and is located in a sky region with a significant lower density of massive galaxies than those found in the center of the Coma, Virgo or Fornax galaxy clusters.

The formation mechanism for these ultra diffuse galaxies is unknown. They have only been found in clusters so far (as is also the case for quenched dwarf galaxies; Gavazzi et al. 2010; Geha et al. 2012) and their Sersic indices, sizes, red colors and round morphologies resemble the most extended dEs (e.g., Geha et al. 2003; Toloba et al. 2014b). Thus, one possibility is that they are the products of environmental effects affecting a progenitor population that falls into a cluster. The structural properties of these diffuse galaxies suggest that they are the extension to larger sizes of dwarf early-type galaxies (dEs; see Figure 8). \DGSATI\ also has a mass in the range of those of dEs (e.g. Toloba et al. 2014b), and its blue off-center kpc-scale over-density closely resembles those found in dEs as well (Lisker et al. 2006).

Tidal effects (including harassment, stripping, stirring, and heating)
are environmental mechanisms produced by gravitational interactions
between galaxies (Moore et al. 1998; Mayer et al. 2001; Mastropietro et
al. 2005).  Stripping and mass loss tend to reduce the sizes of the
affected galaxies, making this an unlikely pathway to forming these ultra
diffuse galaxies.  However, some infalling galaxies could experience
milder tidal heating effects.  In fact, Gnedin (2003) simulated tidal
heating of low-surface-brightness disk galaxies in clusters, and found the
disks could be completely transformed into spheroids while losing few of
their stars and remaining large in size. In combination with ram pressure
stripping to quench the galaxies and remove their gas content (e.g. Lin \&
Faber 1983; Boselli et al. 2008; Boselli et al. 2014; Toloba et al. 2015),
this may be a viable mechanism for the origins of UDGs.  To test this
scenario, more focused simulations are needed, in combination with further
observational information (mainly to determine their total masses through
robust stellar velocity dispersion measurements) about the UDG population. These
models should also explain the presence of apparently isolated UDGs like DGSAT~ I. This would
help to discriminate if these  systems are tidally perturbed versions of a known type of galaxy or
(as previously suggested) a new type of {\it peculiar dwarf} or {\it failed giant} depending upon their
total masses (Beasley et al. 2015). DGSAT I could provide important leverage in discriminating between
the models, owing to its presence at the outskirts -- or beyond -- of a galaxy cluster (Zw 0107+3212: with a projected distance of 1.8 Mpc). One possibility is that it is a ``backsplash'' galaxy that has passed through the center of the cluster (e.g., Gill et al. 2005).

Our detection of this hitherto unknown ultra-diffuse galaxy shows the value of using small aperture (10 to 15$\cm$) apochromatic telescopes with commercial CCD cameras to detect faint, low surface brightness galaxies which cannot be identified through resolved stellar populations or HI surveys. In fact, their wide fields and depths (2--3 magnitudes deeper than the POSS-II survey) make them ideal for uncovering sparse areas of galaxy clusters up to $100\Mpc$ at low surface brightness levels, and check whether a significant population of still undetected, ultra-diffuse galaxies exist. These systems are still a mystery for modern galaxy formation scenarios, and could still remain undetected in large scale optical and radio surveys due to their extremely faint surface brightness and low gas content.


\acknowledgements{ We thank the anonymous referee for helpful comments that improved the manuscript.
 We thank to Martha Haynes for kindly providing an update version of
the Pisces-Perseus supercluster data from the ALFALFA survey plotted in Fig.~5 and
some useful comments.
We also thank Thorsten Lisker and Joachim Janz for providing the Virgo dE 
catalogues and useful comments, Alan McConnachie
for providing the broadband VI photometry of the And II field for
calibration, Sebastian 
Hidalgo for his help to compute the synthetic CMDs for the crowded 
simulations and Carlos Frenk and Till Sawala  for useful comments. DMD devotes this work to 
Ricardo Gonz\'alez Rodr\'\i guez. DMD was partly supported by an 
invited astronomer position at Observatoire de Paris (LERMA) in the 
initial phase of the project. DMD and EKG acknowledge support by Sonderforschungsbereich (SFB) 881 ``The Milky Way System'' of the German Research Foundation (DFB), particularly through subproject A2. TSC acknowledges the support of a National Science Foundation Graduate Research Fellowship. IK and SK were supported
 by the Russian Scientific Foundation grants 13-02-90407, 
13-02-92690 and 14-12-00965. DAF thanks the ARC for financial support via DP130100388. 
JW was supported by NSFC grant 11390372, 11373029 and 11261140641. This 
work was partially supported by the Spanish Ministerio de Econom\'{\i}a y 
Competitividad (MINECO; grant AYA2010-21322-C03-02). ET acknowledges the 
financial support of the Fulbright Program jointly with the Spanish 
Ministry of Education. Based on data collected at Subaru Telescope 
(which is operated by the  National Astronomical Observatory of Japan). 
We acknowledge the very significant cultural role and reverence that the 
summit of Mauna Kea has always had within the indigenous Hawaiian
 community. We are most fortunate to have the opportunity to conduct 
observations from this mountain.}




\begin{figure*}[!htb]
 \centering{\includegraphics[scale=0.7,angle=270]{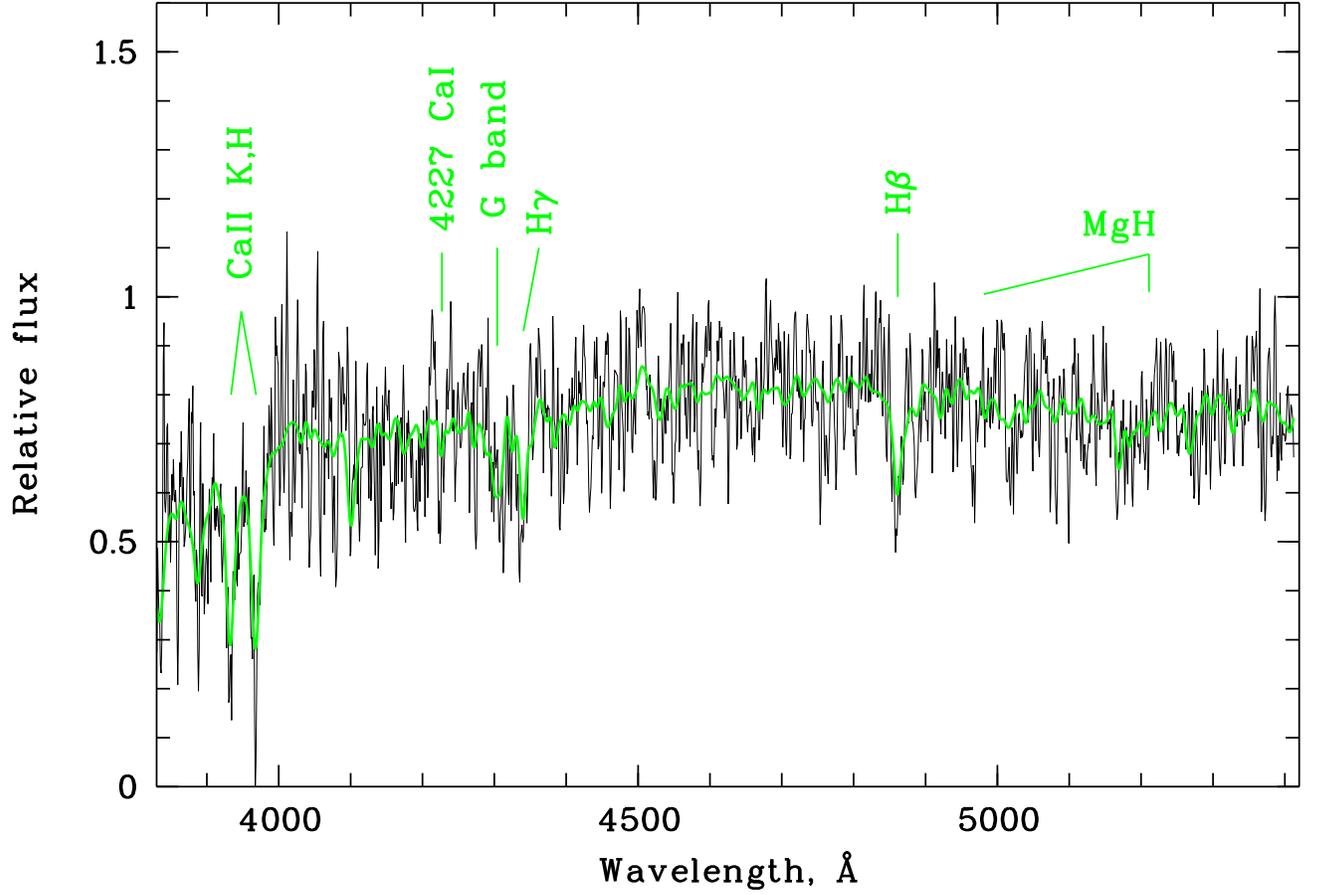}}
\figcaption{The integrated spectrum of the stellar light of \DGSATI\ (black)
 in comparison with a model one (green). The fitting uses
 the Vazdekis et al. (2010) SSP model  composed of intermediate-age metal-rich (age$=1.7 \pm 0.4$\,Gyr, $[{\rm Fe}/{\rm H}]=-0.2 \pm 0.3$\,dex) population and the MILES stellar library. \label{fig3}}
\end{figure*}

\begin{figure*}[!htb]
 \centering{\includegraphics[scale=0.5,angle=270]{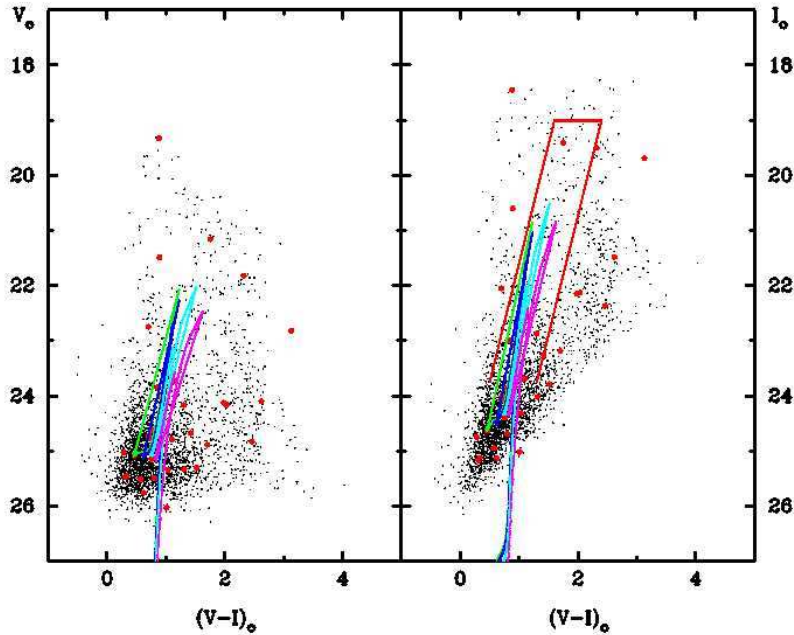}}
\figcaption{Dereddened color-magnitude diagrams of resolved stars in the Subaru/SuprimeCam field
(black dots) and within 40\arcsec (red circles) with sharpness  $|S|\le 1$ and  $\chi< 2.0$ parameters.
The isochrones (Chen et al. 2014) for populations
at a distance modulus for the Andromeda galaxy of 24.6, age 9.8 Gyrs, $[\alpha/\mathrm{Fe}]=+0.2$  and
metallicities $Z$= 1.5$\times$ 10$^{-4}$ (green),
2.4$\times$ 10$^{-4}$ (dark blue),
6.0$\times$ 10$^{-4}$ (pale blue) and 1.4$\times$ 10$^{-3}$ (pink) are indicated, along with the selection
box (red) for resolved RGB stars, which is
very similar to the one used for the detection of the TRGB in fields
around M31 \citep{conn12}.\label{fig4}}
\end{figure*}

\begin{figure*}[!htb]
 \centering{\includegraphics[scale=0.8, angle=0]{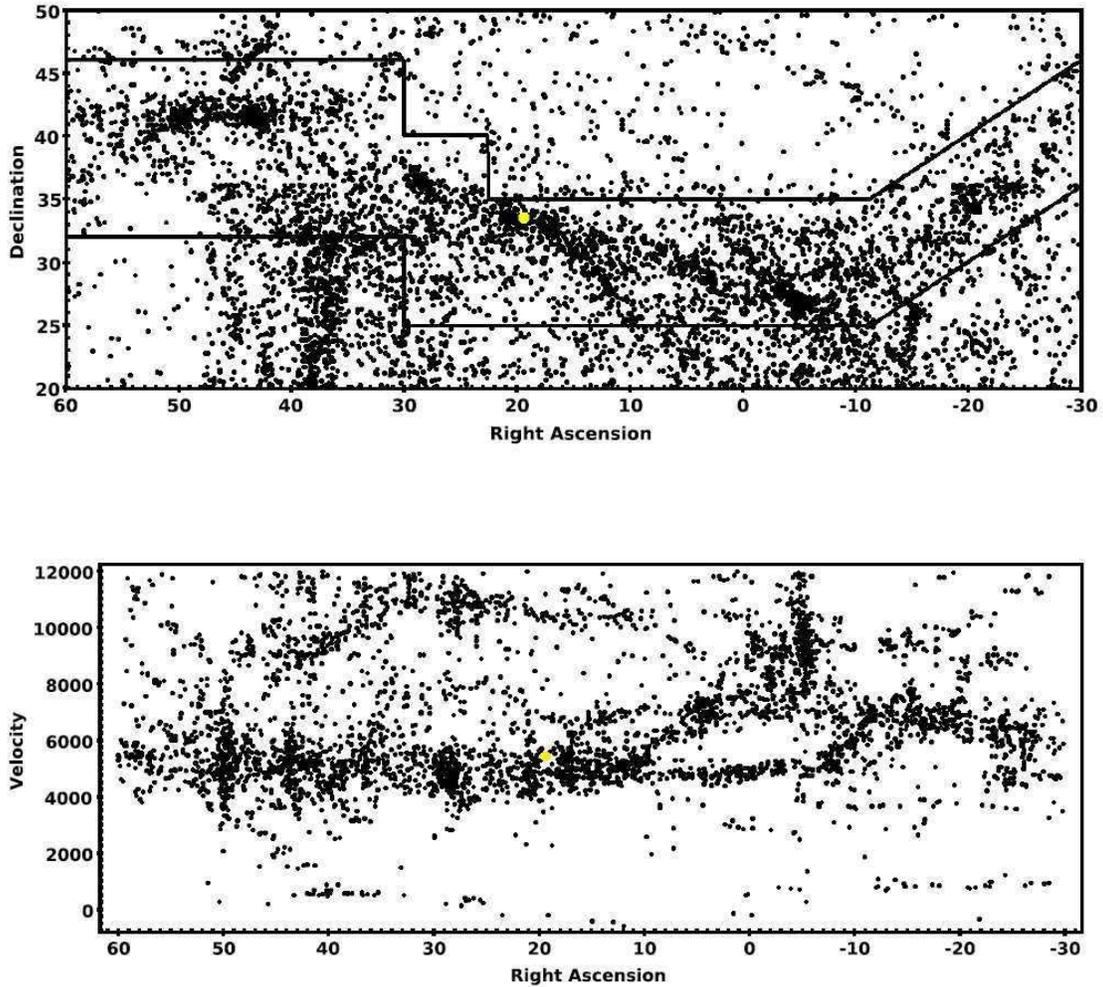}}

\figcaption{Position and measured radial velocity of DGSAT I overplotted on the
available redshift data for the Pisces-Perseus supercluster, following
Wegner, Haynes \& Giovanelli (1993). The redshift dataset includes
measurements from the ALFALFA survey and preliminary ones from
the on-going Arecibo Pisces-Perseus Supercluster survey (APPSS).
Courtesy of M.P. Haynes and the ALFALFA/APPSS team. \label{fig:image}}
\end{figure*}

\begin{figure*}[!htb]
\centering{\includegraphics[width=16cm,angle=270]{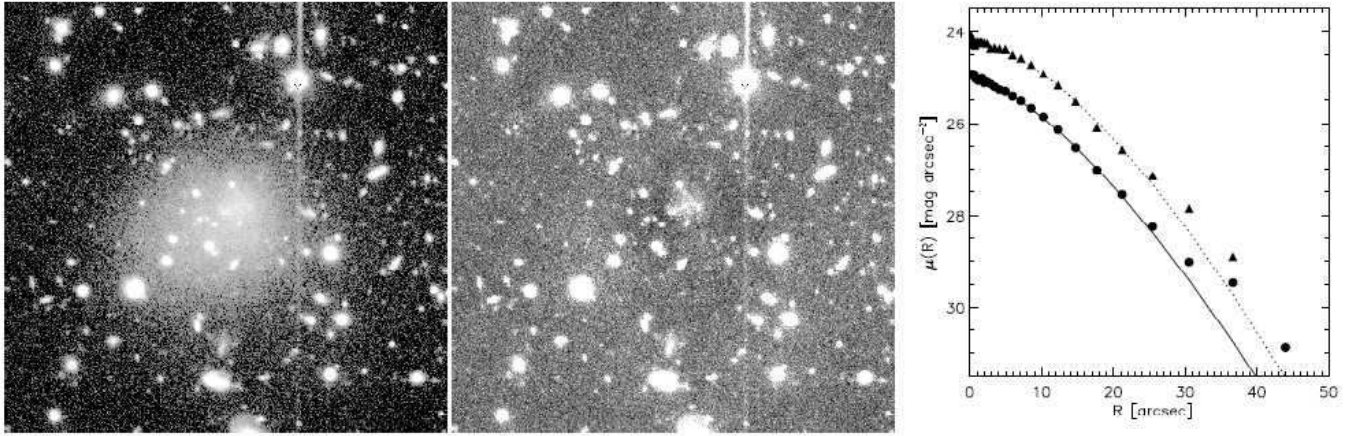}}
\figcaption{GALFIT analysis and surface brightness profile of \DGSATI. {\it Left panel}: cutout from the $I$-band image, 80\arcsec\ on a side (North up, East left).
{\it Middle panel}: best-fit residuals (data minus model) of our adopted \galfit\ model (Model~3, see Section~\ref{subsec:galfit} and Table~1). The data image is displayed on a logarithmic greyscale ranging from $28.5$ (black) to $23.5\magarcsec$ (white); 
a linear scale and a range of $\pm 25\magarcsec$ was used for the residual image. 
Note the lopsidedness of the galaxy, as well as the conspicuous near-central over-density, which is irregularly shaped and flocculent, yet elongated (roughly along the East-West direction), and was masked during the fit. The $V$-band images are very similar and not shown here. {\it Right panel}: Surface brightness profile $\mu(R)$ along the semi-major axis of data and model, displayed respectively as circles and solid line ($V$-band), as well as triangles and dotted line ($I$-band). 
The model agrees with the data well except at $R \gtrsim 25\arcsec$ ($\approx 2R_e$), where it underestimates the surface brightness by $\approx 0.5\mg$. } \label{fig:galfit}
\end{figure*}


\begin{figure*}
\centering
\includegraphics[width=15cm,angle=270]{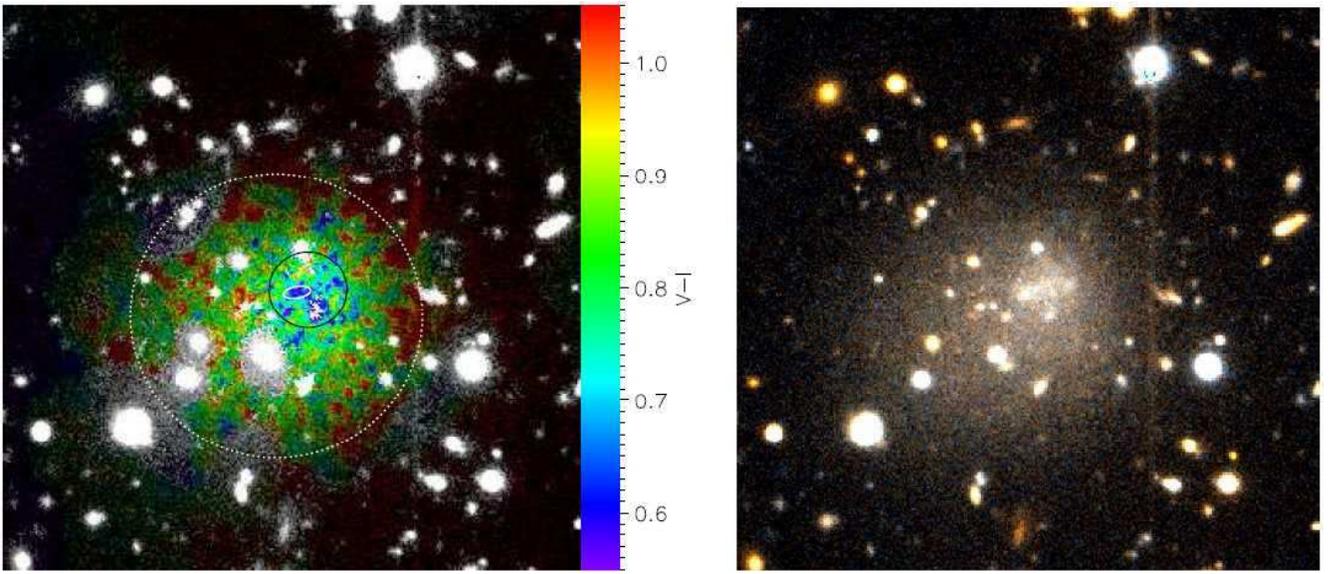}
\figcaption{{\it Left panel:} the $V-I$ color map of \DGSATI, derived by dividing the Subaru $V$- and $I$-band images, after an adaptive boxcar smoothing that ensures a signal-to-noise ratio of $S/N>10$ in the flux ratio. Color values are corrected for Galactic foreground extinction, and pixels that were masked for photometry are displayed on a greyscale instead of color. The over-density slightly offset from the galaxy center stands out by its bluer $V-I$ index. Indeed, measuring the color in two apertures, one large circular annulus with $4<R<15$\arcsec~ from the center (the area between the solid black and dotted white circles), and a small elliptical aperture (solid white) with semiaxes $a$=1\farcsec4, $b$=0\farcsec6, we find $V-I=1.00$ and $0.71$ respectively (see Sec.~\ref{subsec:galfit}). {\it Right panel:} A true-color map of \DGSATI , with red and blue channels proportional to $I$- and $V$-band flux, and the green channel representing the mean of $V$ and $I$-band. Although the color contrast is naturally lower here, the center-offset over-density is recognized as mostly white against the orange-grey of the outer part of the galaxy. Both images are $1'$ on a side and intensity scaled linearly with surface brightness. \label{fig:colormap}}
\end{figure*}

\begin{figure*}[!htb]   
\centering{\includegraphics[scale=1.0,angle=-90]{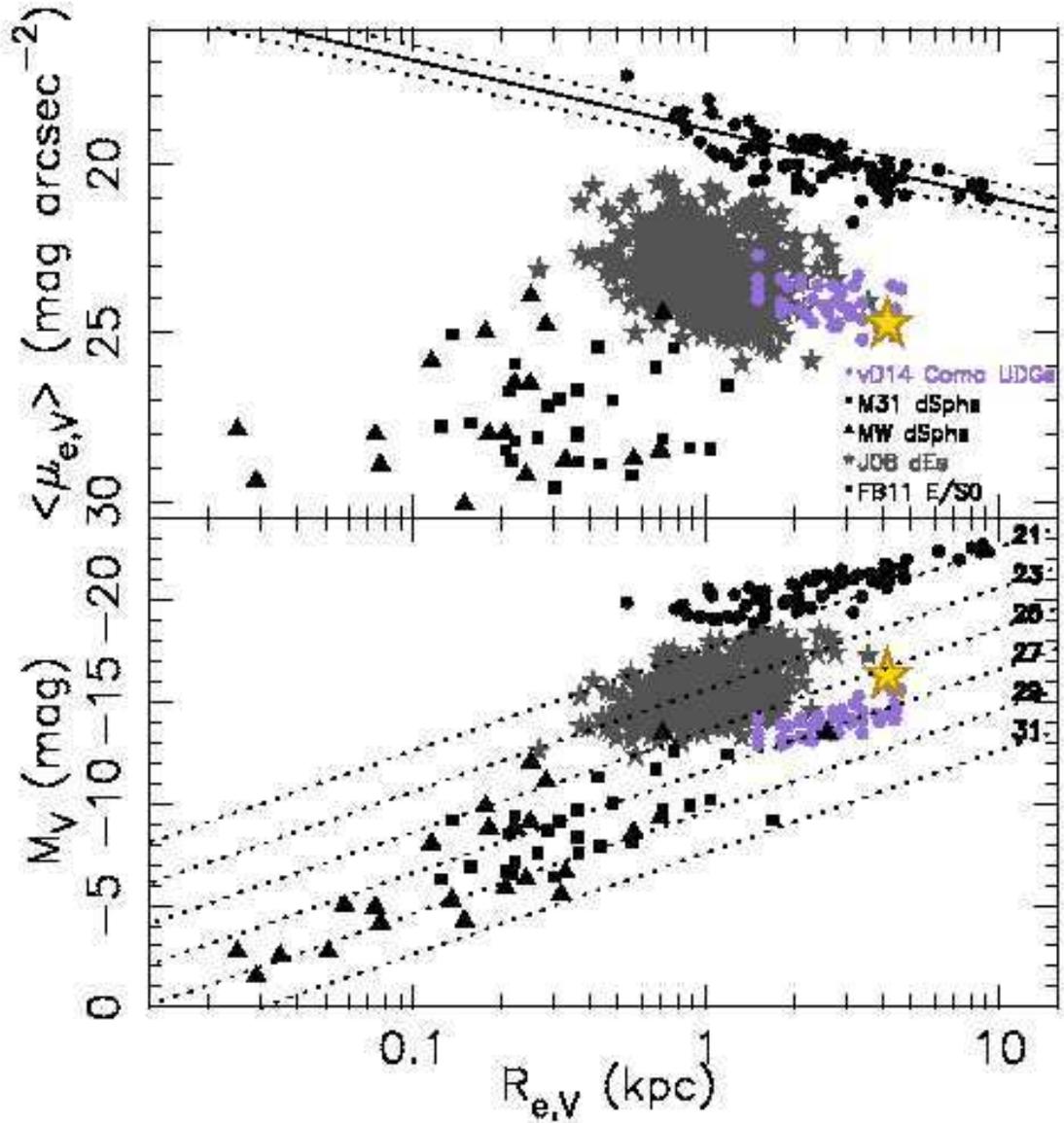}}
\figcaption{Scaling relations for early-type galaxies. {\it Top panel}: surface brightness versus size in $V$-band relation adapted from Toloba et al. (2012). The black dots represent massive elliptical (E) and lenticular (S0) galaxies from Falc\'o n-Barroso et al. (2011). The grey asterisks are Virgo dwarf early-types from Janz \& Lisker (2008; 2009), the black triangles 
and squares are Milky Way and M31 dwarf 
spheroidal galaxies, respectively, from McConnachie et al. (2012).  
The black solid and dotted lines are the best fit
relation and $\pm1\sigma$ scatter for the massive E/S0 galaxies from FB11. 
The yellow star indicate its position, assuming its likely membership in the Pisces-Perseus supercluster with a line-of-sight distance of 78\,Mpc. {\it Bottom panel}: Size-luminosity relation for early-type galaxies in 
the $V$-band. Symbols are as in the top panel. The dotted lines are 
lines of constant surface brightness. Their values in $\magarcsec$ are 
indicated at the right end of those lines\label{fig:scalerels}}
\end{figure*}    



\newpage


\begin{table*}
\centering
\scriptsize
\begin{tabular}{lccccc}
\hline\hline
& band & \multicolumn{2}{c}{Alternative models} & \bf{adopted Model} & modeling uncertainty \\
\cmidrule(lr){3-4}
& & Model 1 & Model 2 & \bf{Model 3} \\
\hline
treatment of blue & & fitted as part & separately modeled by & \bf{masked} \\
central over-density & & of the galaxy & an additional component & \bf{before fitting} \\
\\
 & $V$ & 18.13 & 18.19 & \bf{18.18} & 0.03 \\ 
$m\,[\mathrm{mag}]$ & $I$ & 17.15 & 17.23 & \bf{17.17} & 0.04 \\ 
 & $V-I$ & 0.99 & 0.97 & \bf{1.01} & 0.02 \\ 
 \\
 & $V$ & 24.46 & 24.93 & \bf{24.76} & 0.24 \\ 
$\mu_c\,[\mathrm{mag\,arcsec^{-2}}]$ & $I$ & 23.82 & 24.21 & \bf{23.97} & 0.20 \\ 
 & $V-I$ & 0.65 & 0.72 & \bf{0.81} & 0.08 \\ 
 \\
 & $V$ & 11.3 & 12.2 & \bf{11.7} & 0.5 \\ 
\raisebox{0.5em}[0cm][0cm]{$R_e\,[\mathrm{arcsec}]$} & $I$ & 12.1 & 13.1 & \bf{12.5} & 0.5 \\ 
\\
 & $V$ & 0.80 & 0.62 & \bf{0.68} & 0.09 \\ 
\raisebox{0.5em}[0cm][0cm]{$n$} & $I$ & 0.69 & 0.58 & \bf{0.63} & 0.06 \\ 
\\
 & $V$ & 0.89 & 0.87 & \bf{0.88} & 0.01 \\ 
\raisebox{0.5em}[0cm][0cm]{$b/a$} & $I$ & 0.87 & 0.87 & \bf{0.87} & 0.01 \\ 
\hline
\end{tabular}
\label{Table1}
 \caption{\gfthree\ model photometric properties of \DGSATI\ in our Subaru images, calibrated to the Johnson $V$- and $I$-bands. Rows list (from top to bottom and separately for each band) the extinction-corrected apparent magnitude and color, central surface brightness and central color, effective radius, S\'ersic index, and axis ratio of the 2D-S\'ersic profile. In addition to the adopted model ("Model~3", last column), we show results for two alternative models that differ from the adopted model in the way the central blue over-density is treated. Model~1 was fit with a single \sersic\ and the overdense region left unmasked. Model 2 accounts for this feature by adding a second S\'ersic component. Model~3, which is our preferred model (see also Fig.~\ref{fig:galfit}), was fitted after masking the over-density, but is otherwise the same as the alternative Model~1: a single S\'ersic component with a 1st-order Fourier mode to allow for the lopsidedness. The final column gives the standard deviation between the models as an estimate of the systematic uncertainty, i.e. uncertainty in the choice of model used to parametrize of the light distribution. }
\end{table*}

\begin{table*}
\centering
\scriptsize
\begin{tabular}{lll}
\hline\hline
quantity & notation & value \\
\hline
Right Ascension & RA & $\mathrm{01h\,17m\,35.59s}$ \\
Declination & Dec & $\mathrm{+33^\circ \, 31' \,42 \farcsec 37}$ \\
radial velocity (heliocentric) & $V_h$ & $(5450 \pm 40) \,\mathrm{km\,s^{-1}}$ \\
Hubble distance & $D$ & $(78 \pm 1)\,\mathrm{Mpc}$ \\
\\
apparent magnitude & $m_V$ & $18.18 \pm 0.04$ \\
                                & $m_I$ & $17.17 \pm 0.05$ \\
central surface brightness & $\mu_{c,V}$ & $(24.8 \pm 0.2) \, \mathrm{mag\,arcsec^{-2}}$ \\
                                         & $\mu_{c,I}$ & $(24.0 \pm 0.2) \, \mathrm{mag\,arcsec^{-2}}$ \\
absolute magnitude & $M_V$ & $-16.3 \pm 0.1$ \\ 
                                & $M_I$ & $-17.3 \pm 0.1$ \\ 
luminosity & $L_V$ & $(2.7 \pm 0.2) \times 10^8 \, L_{\odot,V}$ \\ 
                 & $L_I$ & $(3.6 \pm 0.2) \times 10^8 \, L_{\odot,I}$ \\ 
total color & $V-I$ & $1.0 \pm 0.1$ \\
central color & $(V-I)_c$ & $0.8 \pm 0.1 $~($0.71 \pm 0.02$) \\
\\
effective radius & $R_{e,I}$ & $(4.7 \pm 0.2) \,\mathrm{kpc}$ \\
axis ratio & $q_I=(b/a)_I$ & $0.87 \pm 0.01$ \\
S\'ersic index & $n_I$ & $0.6 \pm 0.1$ \\
\\
over-density luminosity & & $\sim 7 \times 10^6 \, L_{\odot,V}$~($\sim 2.5\%$ of \DGSATI) \\
                                    & & $\sim 4 \times 10^6 \, L_{\odot,I}$~($\sim 1.1\%$ of \DGSATI)  \\
over-density effective radius & & $\sim 1.0 \,\mathrm{kpc}$ (both bands, $\sim 22\%$ of \DGSATI) \\
\\                                   
stellar mass-to-light ratio in $I$-band & $M_\star/L_I$  & $1.1\,M_\odot/L_{\odot,I}$ \\ 
stellar mass & $M_\star$ & $4.0 \times 10^8\,M_\odot$ \\ 
gas mass (HI) & $\log M_\hi[M_\odot]$ & $<8.8$ \\ 
\\
$\ha$ flux & $\log F(\ha)\,[{\rm erg}\s^{-1}\cm^{-2}]$ & $< -15.8$ \\
star formation rate & $\log \mathrm{SFR}\,[M_\odot\,{\rm yr^{-1}}]$ & $< -2.6$ \\ 
specific star formation rate & $\log \mathrm{sSFR}\,[{\rm yr^{-1}}]$ & $< -11 $ \\ 
\hline
\end{tabular}
\label{table2}
\caption{Summary of \DGSATI\ properties, assuming our redshift distance of $78\Mpc$. The extinction-corrected apparent magnitudes in the standard (Johnson-Cousins) $V$- and $I$-band, central surface brightness, luminosity, axis ratio and size are the best-fit \gfthree\ 2D-S\'ersic profile constrained by our Subaru images (see Section \ref{subsec:galfit}). $R_e$, $b/a$ and $n$ are given for the $I$-band, but differ from $V$ only at the percent level. For the central color, two values are given: the central magnitude difference of the \sersic\ models in the two bands, and (in brackets) as measured in a small $1\farcsec4 \times 0\farcsec6$ elliptical aperture. The photometric errors give the approximate systematic uncertainty from the choice of parametric model, except for the aperture-based central color where it is based on a Monte-Carlo realization of the aperture geometry. The error in distance and the calibration error of $0.03\mg$ was added in quadrature where applicable, while random errors from pixel noise are negligible and ignored. Due to the unknown systematic uncertainty in $M_\star/L_I$, no errors are given here and for the derived $M_\star$. }
\end{table*}


\clearpage


{}


\end{document}